# Dynamics of correlation-frozen antinodal quasiparticles in superconducting cuprates

F. Cilento,[1]* G. Manzoni,[1,2] A. Sterzi,[1,2] S. Peli,[3] A. Ronchi,[3,4] A. Crepaldi,[5] F. Boschini,[6,7] C. Cacho,[8] R. Chapman,[8] E. Springate,[8] H. Eisaki,[9] M. Greven,[10] M. Berciu,[6,7] A.F. Kemper,[11] A. Damascelli,[6,7] M. Capone,[12] C. Giannetti,[3]* and F. Parmigiani[1,2,13]

[1] *Elettra-Sincrotrone Trieste S.C.p.A., 34149 Basovizza, Italy.*

[2] *Dipartimento di Fisica, Università degli Studi di Trieste, 34127 Trieste, Italy.*

[3] *Interdisciplinary Laboratories for Advanced Materials Physics (ILAMP), Università Cattolica del Sacro Cuore, Brescia I-25121, Italy.*

[4] *Department of Physics and Astronomy, KU Leuven, Celestijnenlaan 200D, 3001 Leuven, Belgium.*

[5] *Institute of Physics, Ecole Polytechnique Fédérale de Lausanne (EPFL), CH-1015 Lausanne, Switzerland.*

[6] *Department of Physics and Astronomy, University of British Columbia, Vancouver, BC V6T 1Z1, Canada.*

[7] *Quantum Matter Institute, University of British Columbia, Vancouver, BC V6T 1Z4, Canada.*

[8] *CLF-Artemis@Rutherford Appleton Laboratory, Harwell Science and Innovation Campus, Didcot, OX11 0QX, UK.*

[9] *Nanoelectronics Research Institute, National Institute of Advanced Industrial Science and Technology, Tsukuba, Ibaraki 305-8568, Japan.*

[10] *School of Physics and Astronomy, University of Minnesota, Minneapolis, Minnesota 55455, USA.*

[11] *Department of Physics, North Carolina State University, Raleigh, NC 27695, USA*

[12] *Scuola Internazionale Superiore di Studi Avanzati (SISSA) and CNR-IOM Democritos National Simulation Center, Via Bonomea 265, 34136 Trieste (Italy).*

[13] *International Faculty, University of Cologne, Albertus-Magnus-Platz, 50923 Cologne, Germany.*

*email: federico.cilento@elettra.eu (F.C.); claudio.giannetti@unicatt.it (C.G.)

**Many puzzling properties of high-$T_c$ superconducting (HTSC) copper oxides have deep roots in the nature of the antinodal quasiparticles, the elementary excitations with wavevector parallel to the Cu-O bonds. These electronic states are most affected by the onset of antiferromagnetic correlations and charge instabilities and they host the maximum of the anisotropic superconducting gap and pseudogap. In this work, we use time-resolved extreme-ultra-violet (EUV) photoemission with proper photon energy (18 eV) and time-resolution (50 fs) to disclose the ultrafast dynamics of the antinodal states in a prototypical HTSC cuprate. After photoinducing a non-thermal charge redistribution within the Cu and O orbitals, we reveal a dramatic momentum-space differentiation of the transient electron dynamics. While the nodal quasi-particle distribution is heated up as in a conventional metal, new quasiparticle states transiently emerge at the antinodes, similarly to what is expected for a photoexcited Mott insulator, where the frozen charges can be released by an impulsive excitation. This transient antinodal metallicity is mapped into the dynamics of the O-2$p$ bands thus directly demonstrating the intertwining between the low- and high-energy scales that is typical of correlated materials. Our results suggest that the correlation-driven freezing of the electrons moving along the Cu-O bonds, analogous to the Mott localization mechanism, constitutes the starting point for any model of high-$T_c$ superconductivity and other exotic phases of HTSC cuprates.**

## Introduction

Many of the exotic phenomena characterizing the phase diagram of HTSC cuprates originate from the interplay between the Coulomb interactions within the Cu-3$d$ orbitals and the tendency of holes to delocalize via the O-2$p$ mediated hopping (*1,2*). The redistribution of the charges within the Cu-O orbitals in doped materials drives the emergence of unconventional phenomena, such as the pseudogap phase (*3,4*), low-temperature charge instabilities (*2,5*) and $d$-wave superconductivity (*6*). A possible framework (*7-10*) to rationalize these phenomena relies on the assumption that, in lightly doped HTSC, the strong on-site repulsion drives the freezing of the electrons moving along the Cu-O bonds. The consequent breakdown of the Fermi surface (*11*), under the form of a suppression of quasiparticle excitations in the antinodal region of the Brillouin zone, prepares the emergence (*12*) of low-temperature instabilities and the opening of the $d$-wave superconducting gap. Within this picture, ultrafast resonant excitation can be used as an "unconventional" control parameter (*13,14*) to transiently modify the electronic occupancy of the Cu-3$d$ and O-2$p$ orbitals and drive the system towards a transient metallic state (*15*) characterized by the partial recovery of antinodal quasiparticle states at momenta ($\pm\pi$,0) and (0,$\pm\pi$).

Led by this idea, we performed high-temporal-resolution EUV angle-resolved photoemission spectroscopy (TR-ARPES) to map the ultrafast evolution of both antinodal quasiparticles and oxygen 2$p$ bands in a prototypical cuprate superconductor (see Figure 1A-C for an overview of the real- and momentum-space structure). Our experiment unveils an unconventional antinodal dynamics, characterized by the transient emergence of quasiparticle states, that is very similar to what expected for a photoexcited Mott insulator. These results, supported by Cluster Dynamical Mean Field Theory (CDMFT) calculations within the single-band Hubbard model, suggest that the breakdown of the antinodal Fermi surface is a direct consequence of the short-range correlations induced by the on-site Coulomb repulsion. The ultrafast dynamics is triggered by near-infrared (1.65 eV) femtosecond pulses that excite the O-2$p$→Cu-3$d$ charge-transfer process in Bi$_2$Sr$_2$Ca$_{0.92}$Y$_{0.08}$Cu$_2$O$_{8+\delta}$ single crystals (Y-Bi2212) close to the optimal hole concentration ($T_c$=96 K).

## Results

### *Ultrafast photoexcitation in multi-band copper oxides*

In order to address the role of the oxygen bands in the photoexcitation process, we start from the one-hole solution (*17*) of the 5-bands model (see Figure 1), obtained from a generalized Emery model after projecting out the double- occupancy of the Cu sites. The spectrum reveals the presence of two different oxygen bands at binding energies larger than 1 eV. The 2$p_\sigma$ band has a maximum at momentum (0,0) and presents the same symmetry of the Cu-3$d_{x^2-y^2}$ states that give rise to the conduction band. In contrast, the 2$p_\pi$ band, which arises from the hybridization of the O-2$p_{x,y}$ orbitals perpendicular to the O-2$p$ ligand orbitals, has almost no overlap with the conduction band (see colors in Fig. 1C) and is observed as a strong photoemission peak (*18,19*) at momentum ($\pi$,$\pi$) and ≈-1.2 eV binding energy (see Figure 1D). These considerations suggest that the near-infrared (1.5 eV photon energy) excitation scheme, which is almost universally adopted in pump-probe experiments (*14,20-25*), is capable of modifying the equilibrium occupancy of the Cu- ($n_d$) and O-bands ($n_\sigma$,$n_\pi$), thus transferring a fraction $\delta n_d$=-($\delta n_\sigma$+$\delta n_\pi$) of charges from the oxygen to the copper orbitals.

### *Antinodal ultrafast dynamics*

The dynamics of $\delta n_d$ over the entire Brillouin zone is tracked by TR-ARPES (see Methods) in which the *s*-polarized probe photons ($\hbar\omega \approx 18$ eV) maintain (*27*) a temporal structure of the order of 50 fs (see Materials and Methods). In Figure 2A we report the band dispersion along the antinodal (AN) direction (see Fig. 1B). Right after the excitation with fluence of 350 µJ/cm$^2$, the differential image, given by the difference between the pumped and unpumped ARPES images, shows an increase (red) of the counts above the Fermi level ($E_F$) accompanied by a decrease below $E_F$. The nature of this non-equilibrium distribution can be better appreciated by plotting (see Figure 2B) the differential energy distribution curves (EDCs) integrated in the momentum region highlighted by the black rectangle in Fig. 2A. The transient antinodal distribution is characterized by a strong electron-hole asymmetry, for the number of electrons above $E_F$ significantly exceeds the number of photoexcited holes below $E_F$. Interestingly, this effect is maximum along the AN cut, while the differential signal along the nodal (N) direction is symmetric at all timescales. In Figure 2C we analyze the dynamics of the antinodal distribution by plotting the EDCs at different pump-probe delays. The EDCs are modeled as the product of a Fermi-Dirac distribution, $f(E,T_{eff})$, at the effective temperature $T_{eff}$ and the spectral function $A(E) = \Sigma_2^2 / \pi[(E-E_F-\Sigma_1)^2 + \Sigma_2^2]$, where $\Sigma = \Sigma_1 + i\Sigma_2$ is the electronic single particle self-energy. The self-energy is phenomenologically modeled by the expression $\Sigma = i\Gamma + \Delta^2/(E-E_F+i\gamma_p)$, that is commonly adopted for gapped quasiparticles (QPs) (*11*), $\Gamma$ being the QP scattering rate, $\Delta$ the electron-hole symmetric pairing gap and $\gamma_p$ a pair-breaking term. At negative delays, the EDCs (grey curve in Fig. 2C) are reproduced assuming an antinodal pseudogap $\Delta_{AN}=40$ meV (*28*) and $T_{eff}=30$ K, which corresponds to the temperature of the sample. In the first hundreds femtoseconds after the excitation (red curves) the EDCs are only partially reproduced by an increase of $T_{eff}$, since they exhibit an excess signal (dashed area) above $E_F$, which is responsible for the asymmetry of the AN differential EDCs, as reported in panel 2B. The most notable result is that the asymmetry of the antinodal EDCs can be perfectly reproduced (see solid line in Fig. 2B) by assuming a net increase of states at the Fermi level, mimicked by adding a transient additional gapless spectral function, $\delta A_{AN}(E)$ (see Supplementary Materials). We stress that a simple gap filling/closing process cannot account for the observed AN asymmetry. In fact, any modification of a pairing gap, when convolved with the experimental resolution larger than the gap size, would give rise to a symmetric signal (see Supplementary Materials), since the increase in the density of states at the Fermi level is compensated by a decrease of the states at the gap edge. In Fig. 2D we report the dynamics of the AN ($0.6 < k < 0.8$ Å$^{-1}$) asymmetry, as obtained by integrating the photoemission intensity in the [-0.7,0.7] eV energy range. The data (yellow squares) show that the electron-hole asymmetry vanishes on a timescale of $\tau_{AN}=110\pm30$ fs. This value is unrelated to the dynamics of $T_{eff}$ (Fig. 2D), which can be estimated from the broadening of the nodal Fermi-Dirac distribution (see Supplementary Materials). This finding suggests that the observed AN increase of states is not the simple result of a temperature-driven broadening of the electronic occupation of states that are intrinsically electron-hole asymmetric (*29,30*).

*The nature of the antinodal states*
Our observations indicate that the redistribution of a fraction of charges $\delta n_d$ within the Cu-3*d* and O-2*p* oxygen bands induces transient new in-gap electronic states at the antinodes. This mechanism is very similar in nature to the collapse of the correlation gap in a photoexcited Mott insulator (*31*), which is determined by the formation of transient quasiparticle states whose spectral weight is compensated at an energy scale of few electronvolts.

The observed dichotomy between the physics of "conventional" nodal quasiparticles and that of Mott-like antinodal excitations is a key consequence of the strong local correlations, as described at the most fundamental level by the single-band Hubbard model. To address the role of onsite correlations in determining the observed antinodal physics, we performed Cluster Dynamical Mean Field Theory (CDMFT) calculations (see Methods), which capture the *k*-space differentiation of the elementary excitations close to the Fermi level. In Fig. 2E we report the antinodal spectral function, convolved with the experimental resolution, at two different temperatures and for effective parameters which reproduce the physics of optimally and over-doped cuprates (see Methods). The calculations evidence a suppression of AN QPs, which is progressively weakened as the energy of the system is raised under the form of a temperature increase. As a consequence, new states appear at the Fermi level, as can be appreciated by plotting the integral of A(*E*) in the [0,0.2] eV range as a function of the temperature (see inset of Fig. 2E). As expected, this effect vanishes when the correlations are weakened by further increasing the hole concentration (see data for the overdoped system in the inset of Fig. 2E), up to the point of completely recovering a closed Fermi surface at all temperatures (*32-34*). Although the single-band Hubbard model cannot account for the details of the photoinduced charge redistribution within the Cu and O orbitals, it shows that the onsite Coulomb repulsion is the fundamental mechanism underlying the *k*-selective suppression of AN QPs for doping concentrations as large as those corresponding to the maximal $T_c$ (optimal doping). Furthermore, it correctly predicts the emergence of additional in-gap antinodal states when energy is provided to the system.

*Ultrafast dynamics of the oxygen states*
A more comprehensive picture of the relation between $\delta n_d$, $\delta n_\sigma$ and $\delta n_\pi$ can be obtained by tracking the transient behavior of the O-2*p* bands after the O-2*p*→Cu-3*d* resonant excitation. In particular, we focus on the dynamics of the non-bonding O-2$p_\pi$ band, which is clearly observed as a strong asymmetric peak at momentum (π,π) (see Fig. 1D). The position of this band is naturally linked to the occupancy of the Cu-3$d_{x^2-y^2}$ and O-2$p_{\pi,\sigma}$ bands. A simple mean-field calculation (see Methods) shows that the photoinduced change $\delta n_d = -(\delta n_\sigma + \delta n_\pi)$ of the orbital populations would induce a shift of the O-2$p_\pi$ band energy given by:

$$\delta \varepsilon_\pi \approx U_{pp}\left(\frac{1}{2}\delta n_\pi + \delta n_\sigma\right) + 2U_{pd}\delta n_d \qquad (1)$$

where $U_{pp}$ and $U_{pd}$ are the Coulomb repulsions within oxygen sites and between the oxygen and copper orbitals, respectively. Eq. 1 suggests that any spatially-homogeneous combination of holes in the O-2$p_\pi$ and O-2$p_\sigma$ bands would give rise to an energy shift $\delta\varepsilon_\pi$. Considering the experimental fluence (see Methods) and the realistic values (*35*) $U_{pp}$=5 eV and $U_{pd}$=2 eV, we obtain, as lower and upper bounds, $\delta\varepsilon_\pi$=-10 meV and +15 meV, corresponding to the cases in which the whole of the photons is absorbed via the creation of homogeneous $\delta n_\sigma$ and $\delta n_\pi$ distributions, respectively.

Figure 3A displays the differential spectra relative to the O-2$p_\pi$ band for different pump-probe delays. In the first tens of femtoseconds we observe a differential signal, which changes sign (positive: red; negative: blue) at the energy corresponding to the binding energy of the O-2$p_\pi$ levels. At longer times, the differential signal is characterized by a homogeneous negative variation. The quantitative analysis of this dynamics is reported in Fig. 3B. The differential EDCs are fitted by modifying the smallest number of parameters in the exponentially broadened gaussian function (see Supplementary Materials) that reproduces the equilibrium peak shape reported in Fig. 1D. In particular, the signal is reproduced over the entire timescale by assuming: i) a long-lived spectral weight variation; ii) an additional transient gaussian broadening, $\delta\gamma$, which adds to an equilibrium broadening of $\gamma \approx 160$ meV, corresponding to the energy resolution of the experiment. Importantly, the position of the O-2$p_\pi$ peak remains always constant during the time evolution, in contrast to what expected for a homogeneous distribution of $\delta n_\sigma$ and $\delta n_\pi$ excitations (see Eq. 1).

In Figure 3D we report the temporal evolution of $\delta\gamma$. The maximum amplitude of the transient broadening roughly corresponds to the energy difference ($\approx 25$ meV) between the energy shifts expected for the homogeneous excitation of the O-2$p_\sigma$ and O-2$p_\pi$ bands. The natural explanation of this result is that the transient increase of the population in the conduction band, $\delta n_d$, is associated with the formation of a spatially inhomogeneous pattern of excess holes involving either the $\pi$ or the $\sigma$ oxygen bands (see Fig. 3E). This situation gives rise to a gaussian distribution of $\delta\varepsilon_\pi$ reflecting the statistical spatial distribution of the possible mixtures of $\delta n_\pi$ and $\delta n_\sigma$ excitations. This picture is supported by the comparison between the dynamics of $\delta\gamma$ and that of the AN excess population, as shown in Figure 3D. The gaussian broadening of the O-2$p_\pi$ peak recovers its resolution-limited value on a timescale identical to the relaxation time of the excess population ($\tau_{AN}$) measured at the antinode and shown in Figure 2.

As a final proof of the relation between the observed broadening and the creation of a non-thermal charge distribution in the Cu-3$d$ and O-2$p$ bands, we repeated the same experiment with a pump excitation characterized by the same absorbed fluence but a photon energy ($\hbar\omega=0.82$ eV) smaller than that necessary to transfer electrons from the O-2$p_{\sigma,\pi}$ bands to the Cu-3$d_{x^2-y^2}$ conduction band. As shown in Figure 3C, no dynamics of the O-2$p_\pi$ is observed in the case of off-resonant excitation, thus ruling out thermal heating effects as the origin of the observed phenomena.

**Discussion**

Our experiment demonstrates that the suppression of quasiparticle states at the antinodes is originated by a correlation-driven breakup of the Fermi surface, as already suggested by DMFT calculations for the single-band Hubbard model (*7,8,10,16*). In this framework, the photo-excitation drives the evolution of AN states from Mott-like gapped excitations to delocalized QPs, as previously suggested by all optical experiments (*15*). The present results directly impact the current knowledge of the physics of cuprate superconductors. Any model for explaining the $d$-wave superconducting pairing, the antinodal pseudogap and the fragility towards charge ordering should build on a ground state in which the strong on-site Coulomb repulsion drives the freezing of antinodal quasiparticles. Finally, although the resonant optical excitation is spatially homogeneous, it likely gives rise to an inhomogeneous pattern of $\delta n_\pi$ and $\delta n_\sigma$ distributions. Whether this is the consequence of the spontaneous and ultrafast segregation of charges in the O-2$p_{\sigma,\pi}$ bands or it reflects an underlying inhomogeneity of the oxygens (*5,36–38*) remains an open fascinating question that requires further investigation.

More generally, the possibility of manipulating the orbital occupancy offers a new way to control the transient properties in correlated and multiorbital materials and to achieve novel functionalities, such as the photoinduced antinodal metallicity shown in this work. The complete reconstruction of the *k*-space dynamics of the conduction band and of the high binding-energy states demonstrated here will constitute the cornerstone for the next generation of experiments based on the novel EUV and XUV ultrafast sources that are currently being developed. Our results also constitute the benchmark for future realistic models of the bandstructure dynamics of multiband correlated materials, in which the orbital occupation can be manipulated on demand by light.

**Materials and Methods**

*Experimental design*

Time-resolved ARPES measurements were performed at the Materials Science end-station at the Artemis facility (Central Laser Facility, Rutherford Appleton Laboratory, UK). The facility is equipped with a 1 kHz Ti:Sapphire amplified laser system delivering ≈30 fs pulses with a central wavelength of 790 nm. The EUV probe photons (in the range 15-40 eV) are produced by high-harmonic generation (HHG) in an argon gas-jet. The *s*-polarized 11$^{th}$ harmonics at ≈18 eV was selected through a time-preserving monochromator (exploiting gratings mounted in the off-plane geometry to preserve the short temporal duration of the pulses). The energy-distribution curves reported in Figs. 1, 2, 3 were fitted by convolving the appropriate fit function with a gaussian function, which accounts for the experimental energy resolution. The gaussian Full-Width-Half-Maximum (FHWM) is 250 meV for the measurements performed close to the Fermi level and 160 meV for the measurements of the oxygen bands. These values correspond to the different settings of the exit slit of the monochromator of the HHG beamline, optimized for obtaining the best compromise between photon flux and energy resolution.

The N, nAN and AN cuts of the Fermi surface reported in Fig. 2 were obtained by rotating the sample around its azimuthal plane. The sample alignment was determined ex-situ by LAUE diffraction and checked in-situ by LEED technique. The pump beam was generated through a high-energy OPA (HE-TOPAS), and tuned to 1510 nm (0.82 eV). For quasi-resonant excitation the second-harmonics of this photon energy was generated in a thin (0.1 mm) phase-matched BBO crystal, thus obtaining a beam at 755 nm (1.65 eV). Both beams were *s*-polarized. The overall temporal resolution in the pump-probe experiments was <50 fs. The density of excited Cu atoms can be calculated considering the pump energy density (25 J/cm$^3$) and the density of Cu atoms (1.3·10$^{-22}$ cm$^{-3}$). The fraction of excited Cu atoms is thus 0.01.

*Bandstructure calculations*

The 5-bands model we used to calculate the spectrum shown in Fig. 1C is explained in Ref. 17. These results are generated with the same variational method used there to calculate the quasiparticle band which hosts the Fermi level upon doping (the highest energy band shown in panel c). While the position and bandwidth of this band were found to depend on the variational space used, specifically on the maximum number of magnons allowed into the quasiparticle cloud, the O bands, especially the $2p_\pi$ one, are very robust and insensitive to such details. This fact further illustrates the lack of hybridization between these bands.

*Dynamical mean field theory calculations*

The Hubbard model has been solved by means of Cluster Dynamical Mean Field Theory (CDMFT) that maps the full lattice model onto a finite small cluster (here a four-site cluster) embedded in an effective medium that is self- consistently determined as in standard mean- field theory. The method therefore fully accounts for the short-range quantum correlations inside the cluster. The calculations use the dynamical cluster approximation prescription (*39*) and the 4* patching of the Brillouin zone introduced in Ref. 16 and have been performed using finite-temperature exact diagonalization (*40*) to solve the self-consistent cluster problem using eight energy levels in the bath as in several previous calculations. The finite-temperature version of the exact diagonalization has been implemented as discussed in Ref. 16 including typically 40 states in the low-temperature expansion of the observables. It is well established that CDMFT calculations with small cluster sizes reproduce the qualitative phase diagram of cuprates at doping smaller than the experimental ones. This discrepancy depends on the details of the calculations (see Ref. 41). For the CDMFT parameters used in this work, the properties corresponding to optimally doped materials are obtained at a nominal hole concentration of $p=0.09$. The properties typical of over-doped materials are obtained at $p=0.2$.

*Mean field estimation of the oxygen orbitals energy*

We consider the interaction piece of the 5-orbital Emery-Hubbard model with on-site and nearest-neighbor interactions in the basis of Cu-3$d_{x^2-y^2}$ and O-2$p_{x,y}$ orbitals:

$$H(\mathbf{r}) = \sum_\alpha \left[ (\varepsilon_d - \mu) n_d^\alpha(\mathbf{r}) + (\varepsilon_\sigma - \mu) n_\sigma^\alpha(\mathbf{r}) + (\varepsilon_\pi - \mu) n_\pi^\alpha(\mathbf{r}) \right] +$$

$$U_{dd} n_d^\uparrow(\mathbf{r}) n_d^\downarrow(\mathbf{r}) + \sum_{\alpha,\alpha'} U_{pd} n_p^\alpha(\mathbf{r}) n_d^{\alpha'}(\mathbf{r}) +$$

$$U_{pp} \left[ n_\sigma^\uparrow(\mathbf{r}) n_\sigma^\downarrow(\mathbf{r}) + n_\pi^\uparrow(\mathbf{r}) n_\pi^\downarrow(\mathbf{r}) \right] +$$

$$\sum_{\alpha,\alpha'} (U_{pp} - 2J_{pp}) n_\sigma^\alpha(\mathbf{r}) n_\pi^{\alpha'}(\mathbf{r})$$

where $n_d$, $n_\sigma$ and $n_\pi$ are the occupations of the Cu-3$d$, O-2$p_\sigma$ and O-2$p_\pi$ orbitals, and $n_p = n_\sigma + n_\pi$. $U_{dd}$ and $U_{pp}$ are the on-site Coulomb interaction for the Cu-3$d$ and O-2$p$ orbitals, and $U_{pd}$ is the intersite interaction. For the interaction $U_{p'p}$ between the $p_\pi$ and $p_\sigma$ orbitals we assume spin-rotation invariance: $U_{p'p} = U_{pp} - 2J_{pp}$.

We perform a mean-field decomposition in the absence of any magnetization ($n^\uparrow = n^\downarrow$), and obtain the following renormalizations of the on-site energy for the $p_\pi$ and $p_\sigma$ orbitals:

$$\delta\varepsilon_\pi \approx U_{pp}\left(\frac{1}{2}\delta n_\pi + \delta n_\sigma\right) + 2U_{pd}\delta n_d - 2J_{pp}\delta n_\sigma$$

$$\delta\varepsilon_\sigma \approx U_{pp}\left(\frac{1}{2}\delta n_\sigma + \delta n_\pi\right) + 2U_{pd}\delta n_d - 2J_{pp}\delta n_\pi$$

## Acknowledgments

*Funding*
C.G. acknowledges support from Università Cattolica del Sacro Cuore through D1, D.2.2 and D.3.1 grants. M.C. and C.G. acknowledge financial support from MIUR through the PRIN 2015 program (Prot. 2015C5SEJJ001). A. C. acknowledges financial support by the SNF. This research was undertaken thanks in part to funding from the Max Planck-UBC Centre for Quantum Materials and the Canada First Research Excellence Fund, Quantum Materials and Future Technologies Program. The work at UBC was supported by the Killam, Alfred P. Sloan, and Natural Sciences and Engineering Research Council of Canada's (NSERC's) Steacie Memorial Fellowships (A.D.); the Alexander von Humboldt Fellowship (A.D.); the Canada Research Chairs Program (A.D.); and the NSERC, Canada Foundation for Innovation (CFI), and CIFAR Quantum Materials. The research leading to these results has received funding from LASERLAB-EUROPE (grant agreement no. 654148, European Union's Horizon 2020 research and innovation programme).

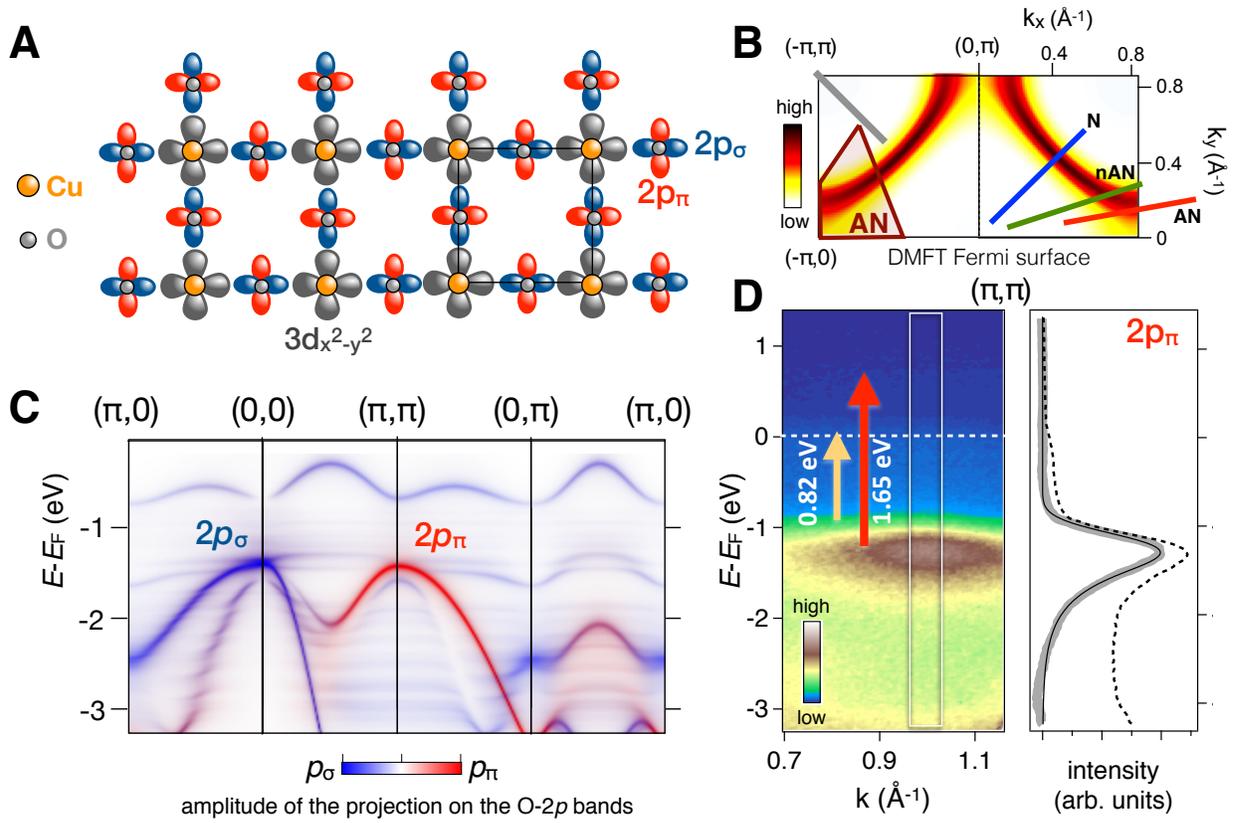

**Fig. 1. Bandstructure of copper oxides. A**) Schematic representation of the Cu-$d_{x^2-y^2}$ and the O-$2p_{x,y}$ orbitals that are included in the five-band model to reproduce the electronic bandstructure. **B**) Upper half of the momentum-space unit cell (Brillouin zone). The Fermi surface is reconstructed by CDMFT (see Methods), for an optimally doped system. The red area indicates the antinodal (AN) region within the 4* patching of the Brillouin zone introduced in Ref. 16 for 2x2 CDMFT. The colored lines represent the nodal (N, blue), nearly antinodal (nAN, green) and antinodal (AN, red) cuts of the Brilloun zone accessible by photoemission with photon energy $\hbar\omega \approx 18$ eV and azimuthal angles $\theta_N=0$, $\theta_{nAN}=27°$ and $\theta_{AN}=35°$. The grey line represents the cut along the $(0,0)$-$(\pi,\pi)$ direction corresponding to the photoemission spectra from the O-$2p_\pi$ band. **C**) Bandstructure from the one-hole solution of the five-band generalized Emery model (see Methods). The color scale indicates the amplitude of the projection of the wavefunctions on the $p_\sigma$ (blue) and $p_\pi$ (red) orbitals. **D**) The left panel displays the photoemission spectrum from the $2p_\pi$ oxygen bands at momentum $(\pi,\pi)$. The color scale (arbitrary units) indicates the photoemission intensity. The arrows indicate the two different pumping schemes used in the experiments. The white rectangle highlights the area of integration for the EDCs reported in the right panel (dashed line). The peak attributed to the $2p_\pi$ band (grey line) is obtained by subtraction of an integral background. The black solid line is the result of the fit of an exponentially-modified gaussian curve to the EDC (see Supplementary Materials).

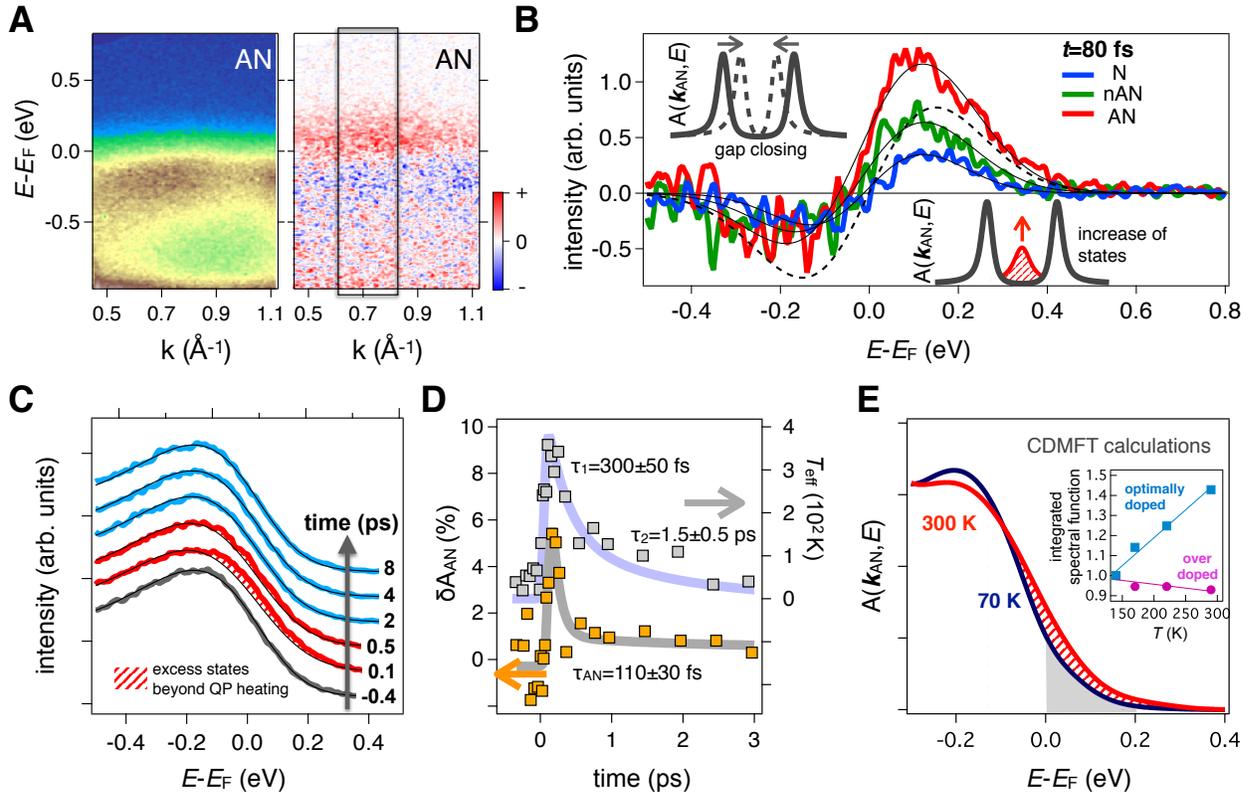

**Fig. 2. Time-resolved photoemission spectra at the Fermi level. A)** Equilibrium (left) and differential (right) anti-nodal (AN) band dispersion. The differential spectrum is obtained as the difference between the pumped and unpumped ARPES images at fixed pump-probe delay (80 fs). The color scale of the differential spectrum highlights positive (red) and negative (blue) photoemission intensity variations. The black rectangle indicates the region of integration for the EDC curves shown in panels B and C. **B)** Differential EDC curves along the N, nAN and AN directions. The black lines are the best fit to the N, nAN and AN differential spectra, obtained by assuming a transient increase of the QP states and of the effective electronic temperature. The dashed line schematizes the symmetric signal expected for a gap closing/filling or a temperature increase. **C)** Non-equilibrium EDC curves at different time delays. The dashed areas show the excess signal, with respect to a simple effective heating, related to the transient increase of states at the Fermi level. The colors highlight three different characteristic temporal regions corresponding to: negative delays (grey trace), short dynamics characterized by the excess antinodal population (red traces), long dynamics characterized by an increase of the electronic effective temperature (blue traces). The black lines show the effective-temperature-increase contribution to the total fit (see Supplementary Materials), which also includes the variation of states at the Fermi level. **D)** The dynamics of the antinodal increase of states (yellow squares, left axis), obtained by integrating the AN spectrum over the momentum-energy area indicated in panel A, is reported. The grey line represents the best fit which contains a single exponential decay with time constant $\tau_{AN}=110\pm30$ fs. The effective temperature increase of the nodal Fermi-Dirac distribution (grey squares, right axis) is well reproduced by a double exponential decay. The extracted timescales are in agreement with published results obtained in similar experimental conditions (*26*). **E)** CMDFT solutions of the single-band Hubbard model for an optimally doped cuprate (see Methods). The lines represent the single-particle spectral functions in the antinodal region (see fig. 1B) at two different temperatures convolved with the experimental resolution. The inset shows the integral of the spectral function in the $0<E-E_F<0.2$ eV energy range (see grey area of the main panel) for optimally and overdoped materials (see Methods). The integrals have been normalized to the values at $T=140$ K.

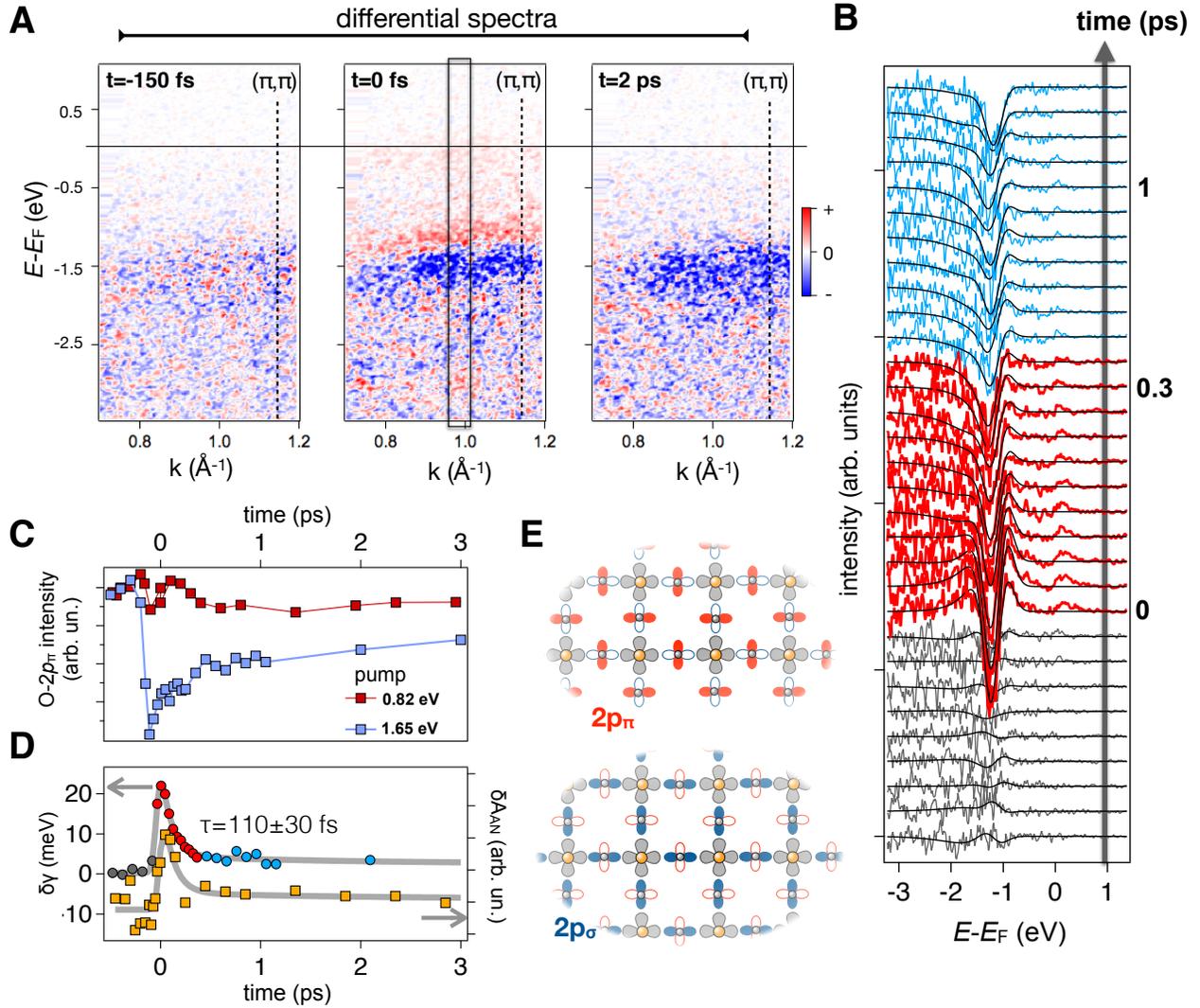

**Fig. 3. Time-resolved photoemission spectra of the oxygen bands. A)** Differential ARPES spectra in the (π,π) momentum region at different time-delays. The color scale highlights positive (red) and negative (blue) photoemission intensity variations. The black rectangle indicates the region of integration for the EDC curves shown in panel B. **B)** Differential EDC curves as a function of the time-delay. The colors highlight three different characteristic temporal regions corresponding to: negative delays (grey traces), short dynamics characterized by a transient broadening of the O-$2p_\pi$ peak (red traces), long dynamics characterized by a long-lived decrease of the O-$2p_\pi$ peak spectral weight. The black lines are the differential fit to the data obtained by assuming both a gaussian broadening and a spectral weight decrease of the O-$2p_\pi$. **C)** Dynamics of the photoemission intensity at -1.2 eV binding energy for the 0.82 eV (red squares) and 1.65 eV (blue squares) pump excitations. **D)** Dynamics of the O-$2p_\pi$ peak broadening (circles, left axis). The colors represent the three different timescales shown in Fig. 3B. For sake of comparison, we show the AN increase of states (yellow squares, right axis), already reported in Fig. 2D. **E)** Cartoon of the inhomogeneous (in real space) excitation pattern of the O-$2p_\pi$ and O-$2p_\sigma$ orbitals at very short timescales (0-300 fs).

## Supplementary Materials

***S1. EDCs at equilibrium and effective temperature dynamics***. Figure S1 shows the nodal (N), near antinodal (nAN) and antinodal (AN) EDCs collected prior to excitation (at pump-probe delay $t$=-400 fs), fitted with the model described in the following. The photoemission intensity is modeled as the product of the spectral function $A(E) = \Sigma_2^2/\pi[(E-E_F-\Sigma_1)^2 + \Sigma_2^2]$, as defined in the main text, and the Fermi-Dirac (FD) distribution $f(E)=(1+\exp((E-E_F)/k_BT))^{-1}$. The resulting expression $A(E)f(E)$ is convolved with a gaussian function accounting for the finite energy resolution of the photoemission experiment. For the EDCs reported in Fig. S1 the base temperature was $T_0$=30 K, and the energy resolution was 250±20 meV. The values of the anisotropic gap $\Delta$ were taken from Ref. 28 and kept fixed at: $\Delta_N$=0, $\Delta_{nAN}$=30 meV, $\Delta_{AN}$=40 meV.

We now turn to the analysis of the dynamics of the effective temperature, as obtained by fitting the nodal EDCs with the model described above. Figure S2 shows the nodal EDCs collected at several pump-probe delays (red lines). The black lines are the fit of the $A(E)f(E)$ model to the data, having as the only free fitting parameter the effective temperature $T_{\text{eff}}(t)$. $T_{\text{eff}}(t)$ is reported in Figure 2D of the main text. The result is $\delta T_{\max}$=320±20 K, being $T_{\text{eff}}(t)=T_0+\delta T(t)$, with the base temperature fixed to $T_0$=30 K. A double-exponential decay was fitted to $T_{\text{eff}}(t)$ to determine its relaxation dynamics (blue solid line). We obtained $\tau_1$=300±50 fs and $\tau_2$=1500±500 fs. The temporal resolution (pump-probe cross-correlation) was better than 50 fs, as determined by the rise-time of $T_{\text{eff}}(t)$.

***S2. Transient modification of photoemission intensity at $E \approx E_F$***. Based on the model described in Section *S1*, we provide graphical illustration of the effect of the change of relevant parameters on the photoemission intensity at $E \approx E_F$, to simulate the evolution after photoexcitation of the Energy Distribution Curves (EDCs). In order to emphasize the effect of each of the parameters on the quantity $A(E)f(E)$ (convolved with the experimental resolution), we report the calculated differential EDCs, i.e., the result of the subtraction of a reference EDC from the EDC with the parameters modified. This representation allows a better comparison with the experimental data, emphasizing the effect that has been photoinduced. Figure S3 summarizes the results of the simulations. A simple heating (red curve), assuming $\delta T$=130 K, leads to a symmetric differential EDC centered about $E=E_F$. A closing of the pairing gap ($\Delta$=40 meV) by $\delta\Delta$=5 meV gives still a symmetric differential EDC, which is offset at higher binding energy with respect to $E_F$ (orange curve). In the case of a filling of the gap $\Delta$, that in the model of Ref. 11 is accounted for by the pair-breaking term $\Gamma_0$, a similar result is obtained, as shown by the yellow curve. In the simulation, we considered $\delta\Gamma_0$=20 meV. Finally, the effect of an increase of the density-of-states (DOS) at $E=E_F$ (green curve) is simulated by adding an additional gapless spectral function, $\delta A(E)$, whose spectral weight corresponds to 3% of the spectral weight of the equilibrium $A(E)$.

The examples reported in Fig. S3 allow drawing important conclusions about the experimental results reported in Fig. 2 of the main manuscript and the effects determining the transient signals observed. While the transient modification of the nodal (N) EDC is fully symmetric and can be mimicked by an effective electronic heating alone ($\delta T_{\max}$=320±20 K, as determined in Section *S1*), the situation near the antinode (nAN) and at the antinode (AN) is different. Here, a non-symmetric differential EDC is observed on a short timescale, while the signal turns to be symmetric on a longer timescale. Hence, the nAN and AN transients have been fitted by introducing both an additional $\delta A_{AN}(E)$ term (on a short timescale) and an effective heating. The best fit to data is obtained with the following values: $\delta T_{max}$=290±30 K and $\delta A_{AN}$=+3.2±0.1% for the nAN EDC and $\delta T_{max}$=320±30 K and $\delta A_{AN}$=+6.0±0.1%.

**S3. Transient broadening of the Oxygen $2p_\pi$ band.** The lineshape of the O-$2p_\pi$ peak, as reported in Fig. 1D of the manuscript, was modeled by the phenomenological function:

$$g(E) = \frac{I}{2\beta}\left\{1 - erf\left[\frac{1}{\sqrt{2}}\left(\frac{E-E_0}{\gamma} - \frac{\gamma}{\beta}\right)\right]\right\} e^{\left[\frac{1}{2}\left(\frac{\gamma}{\beta}\right)^2 - \frac{E-E_0}{\beta}\right]}$$

where $g(E)$ represents an exponentially asymmetric gaussian-broadened peak, which is usually used (*42,43*) to empirically reproduce linewidths whose amplitude is energy dependent (e.g. due to phonon/magnon sidebands or exponentially decaying cross section) and whose width is affected by the experimental energy resolution and by other possible gaussian broadening effects. In this phenomenological function, $I$ represents the intensity of the peak, $E_0$ its binding energy, $\beta$ the exponential asymmetry and $\gamma$ the width (broadening). The exponentially modified gaussian well reproduces the Oxygen $2p_\pi$ peak after an integral background subtraction (thick solid line of Fig. 1D of the manuscript). The fit returned the following parameters: $E_0$=1.2±0.02 eV, $\beta$=0.32±0.01 eV and $\gamma$=0.16±0.005 eV, which corresponds to the energy resolution of the ARPES image reported in Fig. 1D. The modeling of the out-of-equilibrium lineshapes measured at several pump-probe delays (as reported in Fig. 3B of the main manuscript in the differential representation) was performed by fitting $g(E)$ to the curves with two free fitting parameters: the intensity, $I$, and the gaussian broadening of the peak, $\gamma$.

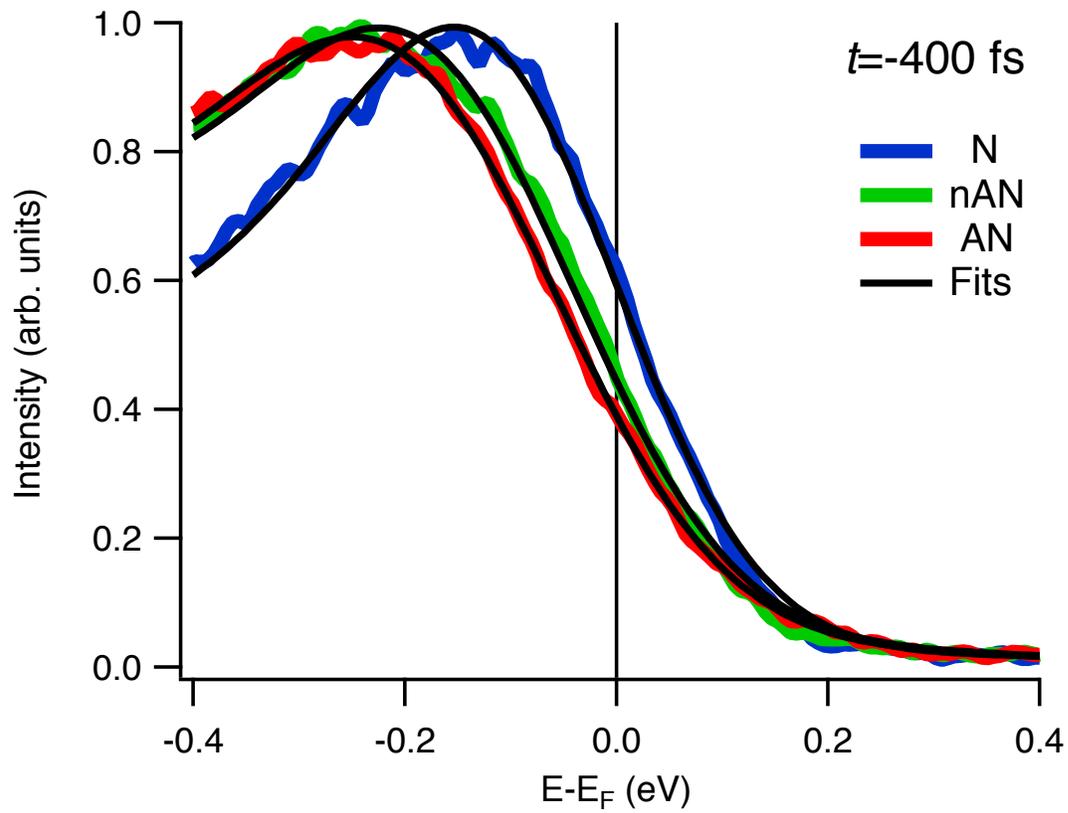

**Fig. S1. Equilibrium energy distribution curves.** The N, nAN and AN EDCs measured at $t=-400$ fs are reported. The solid lines are the fit of the $A(E)f(E)$ expression (convoluted with an experimental resolution of 250±20 meV) to the EDCs.

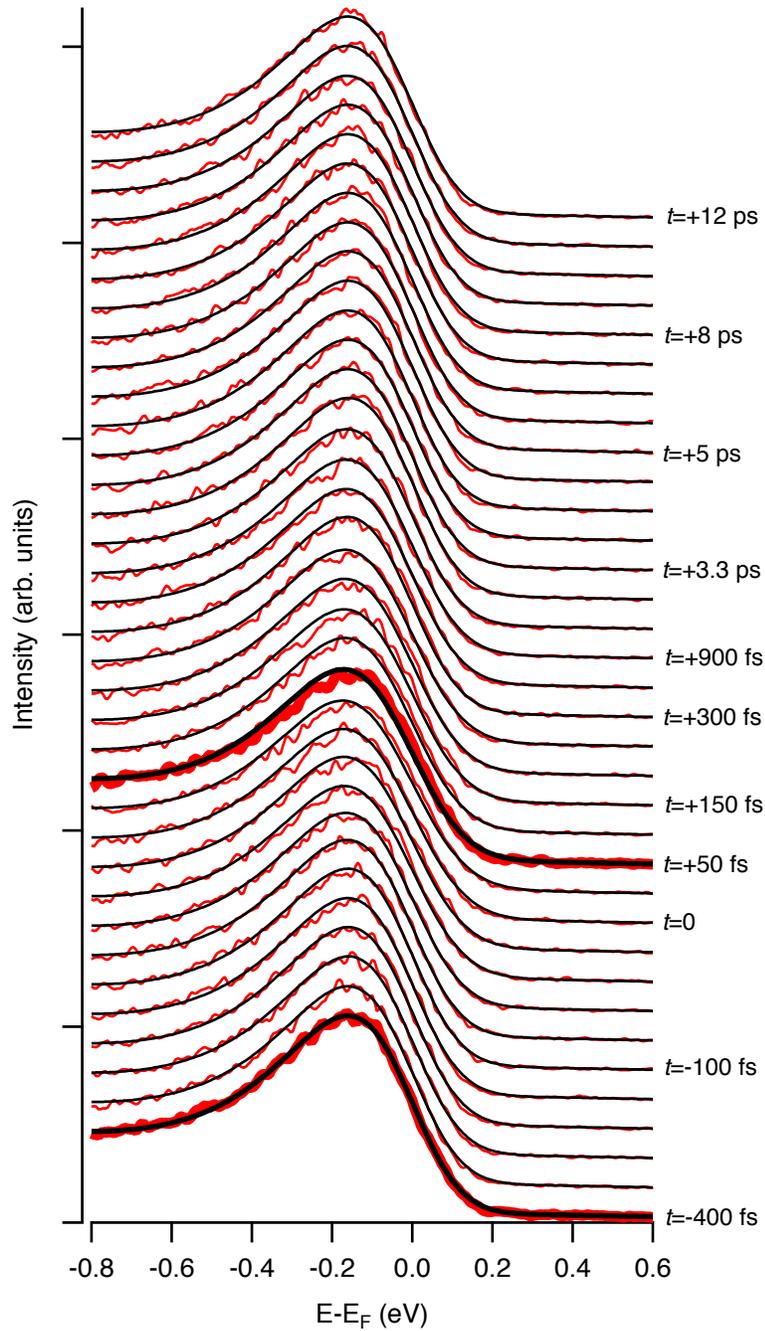

**Fig. S2. Non-equilibrium energy distribution curves.** The nodal EDCs (red curves) are reported for different pump-probe delays, as indicated on the right. The EDCs at $t=-400$ fs and $t=+50$ fs are highlighted. Black lines are the fit to the EDCs, having as the only free fitting parameter the effective electronic temperature $T_e(t)$.

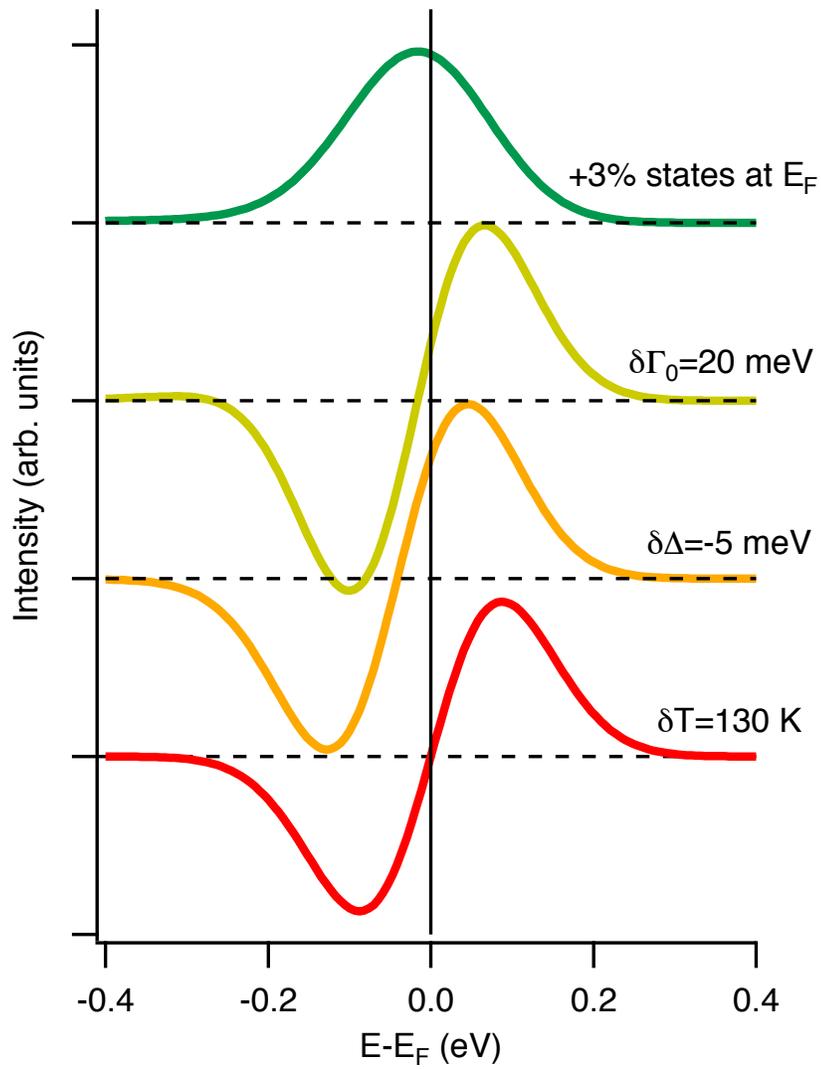

**Fig. S3. Simulations of differential energy distribution curves.** The differential EDCs, as obtained by modifying some of the relevant parameters in the $A(E)f(E)$ function, are shown.